\documentclass[12pt]{article}

\usepackage[usenames,dvipsnames]{color}

\usepackage[centertags]{amsmath}
\usepackage{amsmath}
\usepackage{amsfonts}
\usepackage{amssymb}
\usepackage{amsthm}
\usepackage{newlfont}
\usepackage{amscd}
\usepackage{empheq}
\usepackage{epsfig,graphics,graphicx,amsfonts,psfrag}
\usepackage{verbatim}
\usepackage{eucal}

\newcommand{\NN}{{\mathbb N}}

\newcommand{\CC}{{\mathbb C}}
\newcommand{\beq}{\begin{equation}}
\newcommand{\eeq}{\end{equation}}
\newcommand{\ba}{\begin{array}}
\newcommand{\ea}{\end{array}}
\newcommand{\bea}{\begin{eqnarray}}
\newcommand{\eea}{\end{eqnarray}}
\newcommand{\eps}{{\epsilon}}

\usepackage{graphicx}
\usepackage[usenames,dvipsnames]{color}
\usepackage{transparent} 

\begin{document}

\begin{center}
{\bf The exact rogue wave recurrence in the NLS periodic setting \\via matched asymptotic expansions, for 1 and 2 unstable modes}  

\vskip 10pt

{\it P. G. Grinevich $^{1,3}$ and P. M. Santini $^{2,4}$}

\vskip 10pt

{\it 
$^1$ L.D. Landau Institute for Theoretical Physics, pr. Akademika Semenova 1a, 
Chernogolovka, 142432, Russia, and \\
Lomonosov Moscow State University, Faculty of Mechanics and Mathematics, Russia, 119991, Moscow, GSP-1, 1 Leninskiye Gory, Main Building, and 
Moscow Institute of Physics and Technology, 9 Institutskiy per., Dolgoprudny, Moscow Region, 141700, Russia

\smallskip

$^2$ Dipartimento di Fisica, Universit\`a di Roma "La Sapienza", and \\
Istituto Nazionale di Fisica Nucleare, Sezione di Roma, 
Piazz.le Aldo Moro 2, I-00185 Roma, Italy}

\vskip 10pt

$^{3}$e-mail:  {\tt pgg@landau.ac.ru}\\
$^{4}$e-mail:  {\tt paolo.santini@roma1.infn.it}
\vskip 10pt

{\today}

\begin{abstract}
The focusing Nonlinear Schr\"odinger (NLS) equation is the simplest universal model describing the modulation instability (MI) of quasi monochromatic waves in weakly nonlinear media, the main physical mechanism for the generation of rogue (anomalous) waves (RWs) in Nature. In this paper we investigate the $x$-periodic Cauchy problem for NLS for a generic periodic initial perturbation of the unstable constant background solution, in the case of $N=1,2$ unstable modes. We use matched asymptotic expansion techniques to show that the solution of this problem describes an exact deterministic alternate recurrence of linear and nonlinear stages of MI, and that the nonlinear RW stages are described by the N-breather solution of Akhmediev type, whose parameters, different at each RW appearence, are always given in terms of the initial data through elementary functions. This paper is motivated by a preceeding work of the authors in which a different approach, the finite gap method, was used to investigate periodic Cauchy problems giving rise to RW recurrence. 
\end{abstract}

\end{center}

\textbf{Introduction}. The self-focusing Nonlinear Schr\"odinger (NLS) equation 
\beq\label{NLS}
i u_t +u_{xx}+2 |u|^2 u=0, \ \ u=u(x,t)\in\CC
\eeq
is a universal model in the description of the propagation of a quasi monochromatic wave in a weakly nonlinear medium; in particular, it is relevant in deep water \cite{Zakharov}, in nonlinear optics \cite{Solli,Bortolozzo,PMContiADelRe}, in Langmuir waves in a plasma \cite{Sulem}, and in the theory of attracting Bose-Einstein condensates \cite{Bludov}. It is well-known that its elementary solution
\beq\label{background}
u_0(x,t)=\exp (2it), 
\eeq
describing Stokes waves \cite{Stokes} in a water wave context, a state of constant light intensity in nonlinear optics, and a state of constant boson density in a Bose-Einstein condensate, is unstable under the perturbation of waves with sufficiently large wave length \cite{Talanov,BF,Zakharov,ZakharovOstro,Taniuti,Salasnich}, and this modulation instability (MI) is considered as the main cause for the formation of rogue (anomalous, extreme, freak) waves (RWs) in Nature \cite{HendersonPeregrine,Dysthe,Osborne,KharifPeli1,KharifPeli2,Onorato2}. RWs are transient waves appearing, apparently from nowhere, in all the above physical contexts.

The integrable nature \cite{ZakharovShabat} of the NLS equation allows one to construct a family of exact solutions corresponding to perturbations of the background (\ref{background}), using degenerations of finite-gap solutions \cite{Its,BBEIM,Krichever2,Krichever3} or, more directly, using classical Darboux \cite{Matveev0,Ercolani} - Dressing \cite{ZakharovShabatdress,ZakharovMikha} techniques. Among these basic solutions, we mention the Peregrine soliton \cite{Peregrine}, rationally localized  in $x$ and $t$ over the background (\ref{background}), the so-called Kuznetsov \cite{Kuznetsov} - Ma \cite{Ma} soliton, exponentially localized in space over the background and periodic in time, and the Akhmediev breather \cite{Akhmed1,Akhmed2}. These solutions have also been generalized to the case of multi-soliton solutions, describing their nonlinear interaction, see, f.i., \cite{DGKMatv,Its,Hirota,Akhm6,ZakharovGelash2}. Generalizations of these solutions to the case of integrable multicomponent NLS equations have also been found \cite{BDegaCW,DegaLomb}. 

Concerning the NLS Cauchy problems in which the initial condition consists of a perturbation of the exact background (\ref{background}), if such a perturbation is localized, then slowly modulated periodic oscillations described by the elliptic solution of (\ref{NLS}) play a relevant role in the longtime regime \cite{Biondini1,Biondini2}. If the initial perturbation is $x$-periodic, numerical experiments and qualitative considerations indicate that the solutions of (\ref{NLS}) exhibit instead time recurrence \cite{Yuen1,Yuen2,Yuen3,Akhmed3,Simaeys,Kuznetsov2}, as well as numerically induced chaos \cite{AblowHerbst,AblowSchobHerbst,AblowHHShober}, in which the almost homoclinic solutions of Akhmediev type seem to play a relevant role \cite{CaliniEMcShober,CaliniShober1,CaliniShober2}. There are reports of experiments in which the Peregrine and Akhmediev solitons were observed \cite{CHA_observP,KFFMDGA_observP,Yuen3,Tulin}, but no analytic proof of their relevance within generic Cauchy problems associated with NLS has been given so far, to the best of our knowledge, prior to our work. 

In the paper \cite{GS1} and in this paper we answear some of these basic questions and others, investigating, by two different approaches: the finite gap method \cite{Novikov,Its2,Krichever} and, respectively, matched asymptotic expansion techniques, the NLS equation (\ref{NLS}) on the segment $[0,L]$, with periodic boundary conditions, and we consider, as initial condition, a generic, smooth, periodic, zero average, small perturbation of the background solution (\ref{background}):
\beq\label{Cauchy}
\ba{l}
u(x,0)=1+\eps(x), \ \ \eps(x+L)=\eps(x), \   
||\eps(x)||_{\infty}=\eps \ll 1, \  \int\limits_{0}^L\eps(x)dx =0 .
\ea
\eeq
It is well-known that, in this Cauchy problem, the MI is due to the fact that, expanding the initial perturbation in Fourier components:
\beq\label{Fourier}
\eps(x)=\sum\limits_{j\ge 1}\left(c_j e^{i k_j x}+c_{-j} e^{-i k_j x}\right), \ \ k_j=\frac{2\pi}{L}j , \ \ |c_j |=O(\eps), 
\eeq
and defining $N\in\NN^+$ through the inequalities $\pi L -1 < N < \pi L$, 
the first $N$ modes $\pm k_j,~1\le j \le N$, are unstable, since they give rise to exponentially growing and decaying waves of amplitudes $O(\eps e^{\pm \sigma_j t})$, where the growing factors $\sigma_j$ are defined by
\beq\label{def_ampl}
\sigma_j=k_j\sqrt{4-k^2_j}>0,
\eeq 
while the remaining modes give rise to oscillations of amplitude $O(\eps e^{\pm i \omega_j t})$, where $\omega_j=k_j\sqrt{k^2_j -4}$, and therefore are stable. 

We have in mind the following {\it qualitative recurrence scenario} for finite $N$. The exponentially growing waves become $O(1)$ at times of $O({\sigma_j}^{-1}|\log~\eps|)$, when one enters the nonlinear stage of MI and one expects the generation of a transient, $O(1)$, coherent structure, described by a soliton - like solution of NLS over the unstable background (\ref{background}), the 
so-called RW. Such a RW will have an internal structure, due to the nonlinear interaction between the $N$ unstable Fourier modes, fully described by the integrable NLS theory. Due again to MI, such a coherent RW is expected to be destroyed in a finite time interval, and one enters the third asymptotic stage, characterized, like the first one, by the background plus an $O(\eps)$ perturbation, and described again by the NLS theory linearized around the background. This second linearized stage is expected, due again to MI, to give rise to the formation of a second nonlinear stage of MI. This procedure should iterate forever, in the integrable NLS model, giving rise to the generation of an infinite sequence of RWs. Therefore one is expected to be dealing with the following basic deterministic issues. For a given generic initial condition of the type (\ref{Cauchy}), how to predict: 1)  the ``generation time'' of the first RW; 2) the ``recurrence times'' measuring the time intervals between two consecutive RWs; the analytic form of this deterministic sequence of RWs.

Using the finite gap method, in \cite{GS1} we have studied the case in which the initial perturbation excites a single unstable mode. We distinguished two subcases. In the first subcase in which only the corresponding unstable gap is theoretically open, the solution describes an exact deterministic alternate recurrence of linear and nonlinear stages of MI, and the nonlinear RW stages are described by the 1-breather Akhmediev solution, whose parameters, different at each RW appearence, are always given in terms of the initial data through elementary functions. If the total number $N$ of unstable modes is $>1$, this uniform in $t$ dynamics is sensibly affected by perturbations due to numerics and/or real experiments, provoking $O(1)$ corrections to the result, unless $N=1$. In the second subcase in which more than one unstable gap is open, we obtain the elementary description of the first nonlinear stage of MI, given again by the Akhmediev 1-breather solution, and how perturbations due to numerics and/or real experiments can affect this result. In addition, in \cite{GS3} we have shown that the finite gap recurrence formulas constructed in \cite{GS1} are a good model for describing the numerical (and physical) instabilities of the Akhmediev breather.  

Since, in the first subcase of \cite{GS1}, the solution is given in terms of different elementary functions in different time intervals, obviously matching in the intermediate regions, Matched Asymptotic Expansions (MAEs) techniques are suggested as an alternative approach to the problem and, in this paper, concentrating on the cases of $N=1$ and $N=2$ unstable modes, we show how to construct, via MAEs, the analytic formulas describing the exact RW recurrence, deterministically generated by the generic initial condition (\ref{Cauchy}) and remarkably given in terms of elementary functions. As we shall see, each RW of the sequence is described, in the finite $t$-interval in which it appears, by the $N$-breather solution of Akhmediev type ($x$-periodic, localized in time, and changing the background phase at each appearance). Such solution depends on $(3 N+1)$ real parameters that can be expressed, through elementary functions, in terms of the initial data. Each RW of the sequence is therefore characterized by its own set of $(3N+1)$ parameters, and the simple relation between the set of parameters of two consecutive RWs, generated by the integrable NLS dynamics, is unveiled. If $N=1$, one of course recovers the formulas already derived in \cite{GS1} via the finite gap method. 

These results, presented here for $N=1$ and $N=2$, can be generalized without any conceptual difficulty to the case of an arbitrary number $N$ of unstable modes, and this is one of the research lines we are presently following. We are also exploring, at the same time, the interesting case in which $N\gg 1$ and the situation in which the parameters appearing in the initial condition are random. This last Cauchy problem has been investigated numerically in \cite{AgafontZakharov1}.

We first consider the case in which the initial perturbation (\ref{Cauchy}) excites only the unstable modes. Then, for $|t|\le O(1)$, one can show that:
\beq\label{reg1_N UM}
\ba{l}
u(x,t)=e^{2it}\left(1+\sum\limits_{j=1}^N\left(\frac{|\alpha_j |}{\sin 2\phi_j}e^{\sigma_j t+i\phi_j}\cos[k_j (x-X^+_j)]+ \right.\right. \\
\left.\left. \frac{|\beta_j |}{\sin 2\phi_j}e^{-\sigma_j t-i\phi_j}\cos[k_j (x-X^-_j)] \right)\right) +O(\eps^2) ,
\ea
\eeq
where
\beq\label{def_alpha_beta}
\ba{l}
\alpha_j=\overline{c_j}-e^{2i\phi_j}c_{-j}, \ \ \beta_j=\overline{c_{-j}}-e^{-2i\phi_j}c_j, \\
X^{+}_j=\frac{\arg(\alpha_j)-\phi_j+\pi/2}{k_j}, \ \ X^{-}_j=\frac{-\arg(\beta_j)-\phi_j+\pi/2}{k_j},  \\
\sigma_j=2\sin(2\phi_j), \ \ k_j=2\cos\phi_j \ \ \Leftrightarrow \ \ \phi_j=\arccos(k_j/2), \ \ j=1,\dots,N .
\ea
\eeq
{\it The initial datum splits into exponentially growing and decaying waves, respectively the $\alpha$- and $\beta$-waves, each one carrying half of the information encoded into the initial datum}. At $t=O(|\log\eps |)$, the exponentially growing $\alpha$-waves become $O(1)$ and the solution is described by an exact NLS solution matching with the asymptotic formula  
\beq\label{overlapping1}
u(x,t)\sim e^{2it}\sum\limits_{j=1}^N\left(1+\frac{|\alpha_j |}{\sin 2\phi_j}e^{\sigma_j t+i\phi_j}\cos[k_j (x-X^+_j)]\right),
\eeq
obtained evaluating (\ref{reg1_N UM}) in the intermediate region $1\ll t \ll O(|\log\eps |)$. 

\vskip 5pt
\noindent
{\bf The case $N=1$}. We first specialize our calculation choosing $N=1$. Therefore we are looking, in the nonlinear region $t=O(|\log\eps |)$, for an exact 1-mode, $x$-periodic, transient solution of NLS, matching with (\ref{overlapping1}) for $N=1$ in the overlapping region $1\ll t \ll O(|\log\eps |)$. The natural candidate for such a solution is the well-known Akhmediev 1-breather \cite{Akhmed1,Akhmed2}:
\beq\label{Akhm1}
A_1(x,t;\theta_1,x_1,t_1,\rho)=e^{2it+i\rho}\frac{\cosh[\Sigma_1 (t-t_1)+2i\theta_1 ]+\sin\theta_1 \cos[K_1(x-x_1)]}{\cosh[\Sigma_1 (t-t_1)]-\sin\theta_1 \cos[K_1(x-x_1)]},
\eeq
where
\beq\label{angle_parametriz_2}
K_1=2\cos\theta_1, \ \ \Sigma_1=2\sin(2\theta_1),
\eeq
and $\theta_1,~x_1,~t_1,\rho$ are $4$ arbitrary real parameters to be fixed via MAE. 

It is well-known that this solution is exponentially localized in time over the background $u_0$, changing it by the multiplicative phase factor $e^{4 i\theta_1}$ 
\beq\label{time_asympt_1}
A_1(x,t;\theta_1,x_1,t_1,\rho)\to e^{2it+i(\rho\pm 2\theta_1)}, \ \ \mbox{as} \ \ t\to\pm\infty ,
\eeq
and that its modulus takes its maximum at the point $(x_1,t_1)$, with $|A_1(x_1,t_1;\\ \theta_1,x_1,t_1,\rho)|=1+2\sin\theta_1$ .

Since, in the overlapping region, the solution (\ref{Akhm1}) should match with (\ref{overlapping1}) for $N=1$, it easily follows that $\theta_1=\phi_1$ and, consequently, $K_1=k_1$ and $\Sigma_1=\sigma_1$, where $k_1,~\sigma_1,~\phi_1$ are defined in (\ref{def_alpha_beta}); in addition $x_1=X^+_1$, $\rho=2\phi_1$, and $t_1=T_1(|\alpha_1 |)=O(\sigma^{-1}_1|\log\eps |)$, where
\beq\label{def_T1}
T_1(\zeta)\equiv \frac{1}{\sigma_1}\log\left(\frac{\sigma^2_1}{2 \zeta } \right), \ \ \zeta>0 .
\eeq
Therefore {\it the first RW appears in the finite $t$-interval $|t-T_1(|\alpha_1 |)|\le O(1)$, and is described by the Akhmediev $1$ - breather solution of NLS: 
\beq\label{1RW_1}
u(x,t)=A_1\Big(x,t;\phi_1,X^+_1,T_1(|\alpha_1|),2\phi_1\Big)+O(\eps),
\eeq
whose parameters are expressed in terms of the initial data through elementary functions}. It is important to remark that the first RW contains informations only on half of the initial data (the half encoded in the parameter $\alpha_1$: the $\alpha_1$-wave), and that the modulus of the first RW generated by the initial condition (\ref{Cauchy}),(\ref{Fourier}) acquires its maximum at $t=T_1(|\alpha_1|)$ in the point $x=X^+_1$, mod $L$; and the value of this maximum is
\beq\label{max_1}
|u(X^+_1,T_1(|\alpha_1|))|=1+2\sin\phi_1 <1+\sqrt{3} \sim 2.732.
\eeq
This upper bound, consequence of the formula $\sin\phi_1=\sqrt{1-(\pi/L)^2}, \ \pi <L<2\pi$,  
is obtained when $L\to 2\pi$. We also notice that the position $x=X^+_1$ of the maximum of the RW coincides with the position of the maximum of the growing sinusoidal wave of the linearized theory; this is due to the absence of nonlinear interactions with other unstable modes, if $N=1$. 

To find the relation between two consecutive RWs, we could in principle proceed trying to construct the next asymptotic stage, the second stage of linear MI, matching it with the above first RW stage. But this matching is technically difficult for the following reason. The $\alpha_1$-wave, initially $O(\eps)$, becomes the $O(1)$ RW (\ref{1RW_1}) and then decays exponentially, while the $\beta_1$-wave, also $O(\eps)$ initially, becomes $O(\eps^2)$ during the first nonlinear stage of MI, and then grows exponentially, becoming the main responsible for the generation of the second RW of the sequence (this mechanism is also an important source of instability). From these considerations, in order to be able to get the analytic description on the second RW appearance, one should dig to $O(\eps^2)$ when the first RW appears, to extract informations on the hidden $\beta_1$-wave, and this is technically rather difficult. 
Fortunately this difficulty can be overcome by the following simple trick, consisting in going backward in time from the initial condition (\ref{Cauchy}),(\ref{Fourier}). Indeed formula (\ref{reg1_N UM}) describes the NLS dynamics also for finite, negative times; but, in this case, it is the $\beta$ - wave to be dominant in the asymptotic region $1\ll |t|\ll O(|\log\eps|),~t<0$:
\beq\label{overlapping1b}
u(x,t)\sim e^{2it}\left(1+\frac{|\beta_1 |}{\sin 2\phi_1}\cos[k_1(x-X^-_1)]e^{-\sigma_1 t-i\phi_1} \right).
\eeq
It follows that, in the nonlinear region $|t|=O(\sigma^{-1}_1|\log\eps|),~t<0$, the solution is again described by (\ref{Akhm1}), whose parameters are now fixed matching with (\ref{overlapping1b}). Repeating the above calculations, one obtains that, {\it going backward for negative times, the first RW appears in the time interval $|t+T_1(|\beta_1|)|\le O(1)$ and, in this region, is described again by the Akhmediev $1$-breather solution of NLS, but with different parameters:}
\beq\label{1RW_0}
u(x,t)=A_1\left(x,t;\phi_1,X^-_1,-T_1(|\beta_1|),-2\phi_1 \right)+O(\eps)
\eeq

Comparing the two consecutive RWs (\ref{1RW_0}) and (\ref{1RW_1}) and, in particular, their expressions respectively at $t=-T_1(|\beta_1|)$ and $t=T_1(|\alpha_1|)$:
\beq
\ba{l}
u(x,-T_1(|\beta_1|))=e^{-2iT_1(|\beta_1|)-2i\phi_1}\frac{\cos(2\phi_1 )+\sin\phi_1 \cos[\sigma_1(x-X^-_1)]}{1-\sin\phi_1 \cos[k_1(x-X^-_1)]}+O(\eps),\\
u(x,T_1(|\alpha_1|))=e^{2iT_1(|\alpha_1|)+2i\phi_1}\frac{\cos(2\phi_1 )+\sin\phi_1 \cos[\sigma_1(x-X^+_1)]}{1-\sin\phi_1 \cos[k(x-X^+_1)]}+O(\eps),
\ea
\eeq
we see  that the two functions of $x$ coincide, at the leading order, up to an overall multiplicative phase factor and a global shift in the $x$ direction:
\beq
u(x,T_1(|\alpha_1|))=e^{2iT_p +4i\phi_1}u(x-\Delta_1^{(x)},-T_1(|\beta_1|))+O(\eps)
\eeq
where
\beq\label{period}
\ba{l}
T_p=T_1(|\alpha_1|)+T_1(|\beta_1|)=\frac{2}{\sigma_1}\log\left( \frac{\sigma^2_1}{2\sqrt{|\alpha_1\beta_1 |}} \right), \\
\Delta_1^{(x)}=X^+_1-X^-_1=\frac{\arg(\alpha_1\beta_1)}{k_1}.
\ea
\eeq

We conclude that {\it the Cauchy problem (\ref{Cauchy}) gives rise to an infinite sequence of RWs, and the $n^{th}$ RW of the sequence ($n\ge 1$) is described, in the time interval $|t-T_1(\alpha_1)-(n-1)T|\le O(1)$, by the analytic deterministic formula:
\beq\label{RWn_1UM}
\ba{l}
u(x,t)=A_1\Big(x,t;\phi_1,x_1^{(n)},t_1^{(n)},\rho^{(n)} \Big)+O(\eps), \ \ n\ge 1 ,
\ea
\eeq
where
\beq\label{parameters_1n}
\ba{l}
x_1^{(n)}=X^+_1+(n-1)\Delta^{(x)}_1, \ \ t_1^{(n)}=T_1(|\alpha_1|)+(n-1)T_p, \\ \rho^{(n)}=2\phi_1+(n-1)4\phi_1 ,
\ea
\eeq
in terms of the initial data} (see Figures 1). 

\begin{center}
\includegraphics[width=9.6cm]{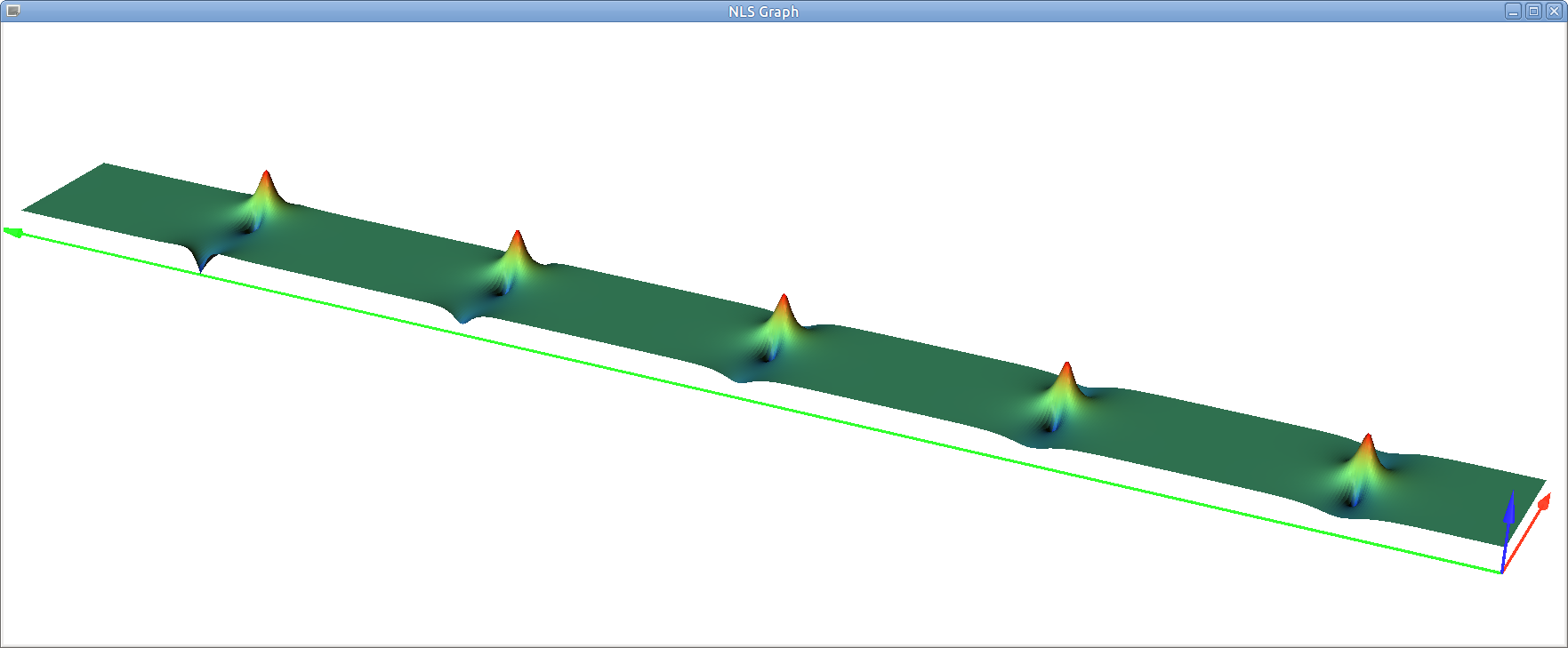} \ \ \includegraphics[width=3.2cm]{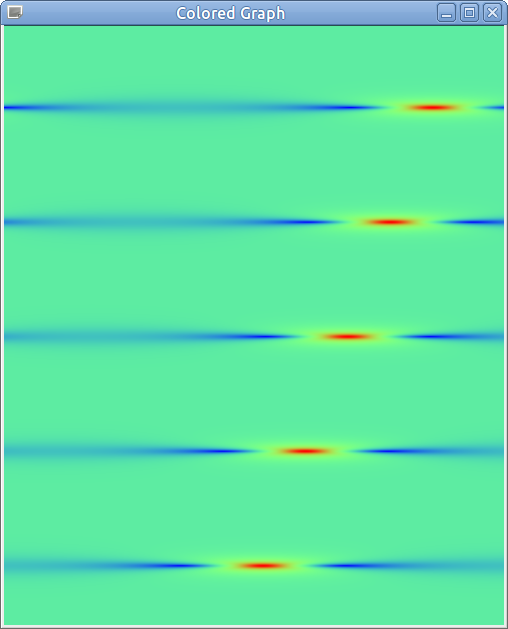}
\end{center}
Figures 1. The 3D and level plots of $|u(x,t)|$ describing the RW sequence, obtained through the numerical integration of NLS via the Split Step Fourier Method (SSFM) \cite{Agrawal,SSFM1,SSFM2}. Here $L=6$ ($N=1$), with $c_1 = \eps/2,~c_{-1} = \eps (0.3-0.4 i)/2,  ~\epsilon = 10^{-4}$, and the short axis is the $x$-axis, with $x\in [-L/2,L/2]$. The numerical output is in perfect agreement with the theoretical predictions.
\vskip 5pt

Keeping in mind the asymptotic properties (\ref{time_asympt_1}) of the Akhmediev breather, one can easily construct the following {\it uniform in 
space-time representation of the solution of the Cauchy problem, in terms of elementary functions, describing the first $n$ RW appearances, 
for $0\le t\le t_1^{(n)}+O(1)$: 
\beq\label{unif_sol_Cauchy_1}
\ba{l}
u(x,t)=\sum\limits_{m=0}^n A_1\Big(x,t;\phi_1,x_1^{(m)},t_1^{(m)},\rho^{(m)} \Big) -\frac{1-e^{4in\phi_1}}{1-e^{4i\phi_1}}e^{2it}, \ \ x\in [0,L], 
\ea
\eeq
where the parameters $x_1^{(m)},~ t_1^{(m)},~\rho^{(m)},~m\ge 0$, are defined in (\ref{parameters_1n}). It turns out that this representation is accurate up to $O(\eps^2|\log\eps|)$ in the regions describing the linear stages of modulation instability, and up to $O(\eps)$ in the regions describing the nonlinear stages of modulation instability (the sequence of RWs).
}

We first remark that each RW of the sequence changes the background exactly by the factor $e^{4i\phi_1}$ and, due to (\ref{max_1}), the upper bound for the amplitude of each RW of the sequence is $(1+\sqrt{3})\sim 2.732$ times the background amplitude.

We also remark that the RW dynamics is periodic also in time, with the period $T_p$ given in (\ref{period}), up to the above overall multiplicative phase factor and the global phase shift in the $x$ direction (see Figures 1). Actually, using the finite gap method, we have been able to show in \cite{GS1} that such a time periodicity is extended also to the time regions in which the solution is described by a small perturbation of the background, through the formula: $u(x,t+T_p)=e^{2iT_p+4i\phi_1}u(x-\Delta_1^{(x)},t)+O(\eps^2|\log\eps |)$, $t\ge 0$.

At last, if the initial condition is the generic one (\ref{Cauchy}),(\ref{Fourier}) for $N=1$, then the solution (\ref{reg1_N UM}) contains also $O(\eps)$ oscillations in the region $|t|\le O(1)$. But its behavior in the overlapping region $1\ll |t|\ll O({\sigma_1}^{-1}|\log\eps|)$, for $t>0$ and $t<0$ is still given by equations (\ref{overlapping1}) and (\ref{overlapping1b}), and the matching is not affected at the leading $O(1)$ . Therefore the sequence of RWs is still described by equations (\ref{RWn_1UM}), (\ref{parameters_1n}), and the differences between the two Cauchy problems are hidden in the $O(\eps)$ corrections. As far as the $O(1)$ RW recurrence is concerned, {\it only the part of the initial perturbation $\eps(x)$ exciting the first unstable mode is relevant}.

The above formulas, first discovered in \cite{GS1} using the finite gap method, have been obtained here using a simpler mathematical tool: matched asymptotic expansions techniques.

\vskip 5pt
\noindent
\textbf{The case $N=2$}. If $N=2$, corresponding to $2\pi<L<3\pi$, and we first choose the simplest nontrivial initial condition in which one excites just the two unstable modes: $u(x,0)=1+\sum\limits_{j=1}^2(c_j e^{i k_j x}+c_{-j} e^{-i k_j x})$, where $c_j,~c_{-j},~j=1,2$ are arbitrary $O(\eps)$ complex parameters and the wave numbers $k_j,~j=1,2$ are defined in (\ref{def_ampl}). This initial condition evolves, for finite times, as in (\ref{reg1_N UM}) with $N=2$, and, in the intermediate region $1\ll t \ll O(|\log\eps |)$, behaves as:
\beq\label{overlapping2}
u(x,t)\sim e^{2it}\left(1+\sum\limits_{j=1}^2\left(\frac{|\alpha_j |}{\sin 2\phi_j}e^{\sigma_j t+i\phi_j}\cos[k_j (x-X^+_j)]\right)\right).
\eeq
This formula must be matched with an exact $2$-mode, $x$-periodic, transient solution of NLS appearing in the nonlinear region $t=O(|\log\eps |)$. Now the natural candidate for such a solution is the $2$-breather solution of Akhmediev type over the background $u_0$ \cite{Akhm6},\cite{CaliniShober1},\cite{ZakharovGelash2}:
\beq\label{Akhm2}
A_2(x,t;\theta_1,\theta_2,x_1,x_2,t_1,t_2,\rho)=e^{2it+i\rho}\frac{N(x,t)}{D(x,t)},
\eeq
where
\beq\label{defN}
\ba{l}
N(x,t)=\cosh[\Sigma_1 (t-t_1)+\Sigma_2 (t-t_2)+2i(\theta_1+\theta_2)]\\
+(a_{12}(\theta_1,\theta_2))^2\cosh[\Sigma_1 (t-t_1)-\Sigma_2 (t-t_2)+2i(\theta_1-\theta_2)]\\
+2a_{12}(\theta_1,\theta_2)\Big\{\sin\theta_1 \cosh[\Sigma_2 (t-t_2)+2i\theta_2]\cos[K_1 (x-x_1)] \\ 
+\sin\theta_2 \cosh[\Sigma_1 (t-t_1)+2i\theta_1]\cos[K_2 (x-x_2)]\Big\} \\
+b^-_{12}(\theta_1,\theta_2)\cos[K_1(x-x_1)+K_2(x-x_2)]\\
+b^+_{12}(\theta_1,\theta_2)\cos[K_1(x-x_1)-K_2(x-x_2)],
\ea
\eeq
\beq\label{defD}
\ba{l}
D(x,t)=\cosh[\Sigma_1 (t-t_1)+\Sigma_2 (t-t_2)]\\
+(a_{12}(\theta_1,\theta_2))^2\cosh[\Sigma_1 (t-t_1)-\Sigma_2 (t-t_2)]\\
-2a_{12}(\theta_1,\theta_2)\Big\{ \sin\theta_1 \cosh[\Sigma_2 (t-t_2)]\cos[K_1 (x-x_1)] \\ 
+\sin\theta_2 \cosh[\Sigma_1 (t-t_1)]\cos[K_2 (x-x_2)]\Big\}\\
+b^-_{12}(\theta_1,\theta_2)\cos[K_1(x-x_1)+K_2(x-x_2)]\\
+b^+_{12}(\theta_1,\theta_2)\cos[K_1(x-x_1)-K_2(x-x_2)],
\ea
\eeq
\beq
\ba{l}
K_j=2\cos\theta_j, \ \Sigma_j=2\sin 2\theta_j, \ j=1,2, \\
~~ \\
a_{12}(\theta_1,\theta_2)=\frac{\sin(\theta_1+\theta_2)}{\sin(\theta_1-\theta_2)}, \ 
b^{\pm}_{12}(\theta_1,\theta_2)=\frac{\sin\theta_1 \sin\theta_2 (\cos\theta_1\pm\cos\theta_2)^2}{\sin^2(\theta_1-\theta_2)},
\ea
\eeq
and the arbitrary $7$ real parameters $\rho,~\theta_j,~x_j,~t_j,~j=1,2$ have to be fixed by matching. The solution (\ref{Akhm2}) changes the background by the phase factor $e^{4i(\theta_1+\theta_2)}$:  
\beq\label{time_asympt_2}
A_2(x,t;\theta_1,\theta_2,x_1,x_2,t_1,t_2,\rho)\to e^{2it+i[\rho\pm 2(\theta_1+\theta_2)]}, \ \ \mbox{as} \ \ t\to\pm\infty .
\eeq

Since (\ref{Akhm2}) has to be matched with (\ref{overlapping2}), we immediately infer, as before, that $\theta_j=\phi_j$ and, consequently, that $K_j=k_j,~\Sigma_j=\sigma_j$, $j=1,2$; in addition $\rho=2(\phi_1+\phi_2)$, $x_j=X^+_j$ and $t_j=T_j(|\alpha_j|), \ j=1,2$, where now the two times $T_j,~j=1,2$ depend on the nonlinear interaction factor $a_{12}(\phi_1,\phi_2)$:
\beq
T_j(\zeta)\equiv \frac{1}{\sigma_j}\log\left( \frac{a_{12}(\phi_1,\phi_2)~\sigma^2_j}{2 \zeta} \right), \ \ \zeta>0, \ \ j=1,2.
\eeq
Therefore {\it the first RW appears in the time interval $|t-(T_1(|\alpha_1|)+T_2(|\alpha_2|))/2)|\le O(1)$ and, in this region, it is described by the $2$-breather solution of Akhmediev type:}
\beq\label{2RW_1}
u(x,t)=A_2\left(x,t;\phi_1,\phi_2,X^+_1,X^+_2,T_1(|\alpha_1|),T_2(|\alpha_2|),2(\phi_1+\phi_2)\right)+O(\eps).
\eeq

To find the relation between two consecutive RWs, it is convenient, as before, to use the same trick used for $N=1$, going backward in time from the initial condition. Indeed formula (\ref{reg1_N UM})  
describes the NLS dynamics also for finite, negative times; but, in this case, the $\beta$ - waves are dominant in the asymptotic region $1\ll |t|\ll O(|\log\eps|),~t<0$:
\beq\label{overlapping2b}
u(x,t)\sim e^{2it}\left(1+\sum\limits_{j=1}^2\frac{|\beta_j |}{\sin 2\phi_j}\cos[k_j(x-X^-_j)]e^{-\sigma_j t-i\phi_j} \right).
\eeq
Therefore, for $|t|=O(|\log\eps|),~t<0$, the solution is again described by (\ref{Akhm2}), whose parameters are now fixed matching with (\ref{overlapping2b}). 

Repeating the above simple calculations, one obtains that, {\it at negative times, the first RW appears in the time interval $|t+(T_1(|\beta_1|)+T_2(|\beta_2|))/2)|\le O(1)$ and, in this region, is described again by the $2$-breather solution of NLS, but with different parameters:}
\beq\label{2RW_0}
u(x,t)=A_2\left(x,t;\phi_1,\phi_2,X^-_1,X^-_2,-T_1(|\beta_1|),-T_2(|\beta_2|),-2(\phi_1+\phi_2)\right)+O(\eps).
\eeq
Comparing the two consecutive RWs (\ref{2RW_0}) and (\ref{2RW_1}) one infers the following simple rule: {\it the NLS dynamics generates a sequence of RWs described by the $2$-breather solution of NLS, and if $\{ x^{(n)}_1,x^{(n)}_2,t^{(n)}_1,t^{(n)}_2,\rho^{(n)}\}$ are the parameters associated with the $n^{th}$ RW, the parameters $\{x^{(n+1)}_1,x^{(n+1)}_2,t^{(n+1)}_1,t^{(n+1)}_2,\rho^{(n+1)}\}$ associated with the $(n+1)^{th}$ RW are given by the simple formulae:
\beq
x^{(n+1)}_j=x^{(n)}_j+\Delta^{(x)}_j, \ t^{(n+1)}_j=t^{(n)}_j+\Delta^{(t)}_j, \ \rho^{(n+1)}=\rho^{(n)}+\Delta^{(\rho)},
\eeq
where
\beq
\ba{l}
\Delta^{(\rho)}=4(\phi_1+\phi_2), \ \ \ \ \Delta^{(x)}_j=X^+_j-X^-_j=\frac{\arg(\alpha_j\beta_j)}{k_j}, \\ 
\Delta^{(t)}_j=T_j(|\alpha_j|)+T_j(|\beta_j|)=
\frac{2}{\sigma_j}\log\left(\frac{a_{12}(\phi_1,\phi_2)~\sigma^2_j}{2\sqrt{|\alpha_j\beta_j|} } \right), \ \ j=1,2,
\ea
\eeq
and the $n^{th}$ RW of the sequence ($n\ge 1$) is described by the analytic deterministic formula:
\beq\label{sequence2RW}
\ba{l}
u(x,t)=A_2\Big(x,t;\phi_1,\phi_2,x_1^{(n)},x_2^{(n)},t_1^{(n)},t_2^{(n)},\rho^{(n)}\Big)+O(\eps), \ \ n\ge 1 ,
\ea
\eeq
where
\beq\label{sequence_param}
\ba{l}
x_j^{(n)}=X^+_j+(n-1)\Delta^{(x)}_j, \ \ t_j^{(n)}=T_j(|\alpha_j|)+(n-1)\Delta^{(t)}_j, \ j=1,2, \\
\rho^{(n)}=2(\phi_1+\phi_2)+(n-1)\Delta^{(\rho)}
\ea
\eeq
in terms of the initial data} (see Figures 2). 
\begin{center}
\includegraphics[width=8.3cm]{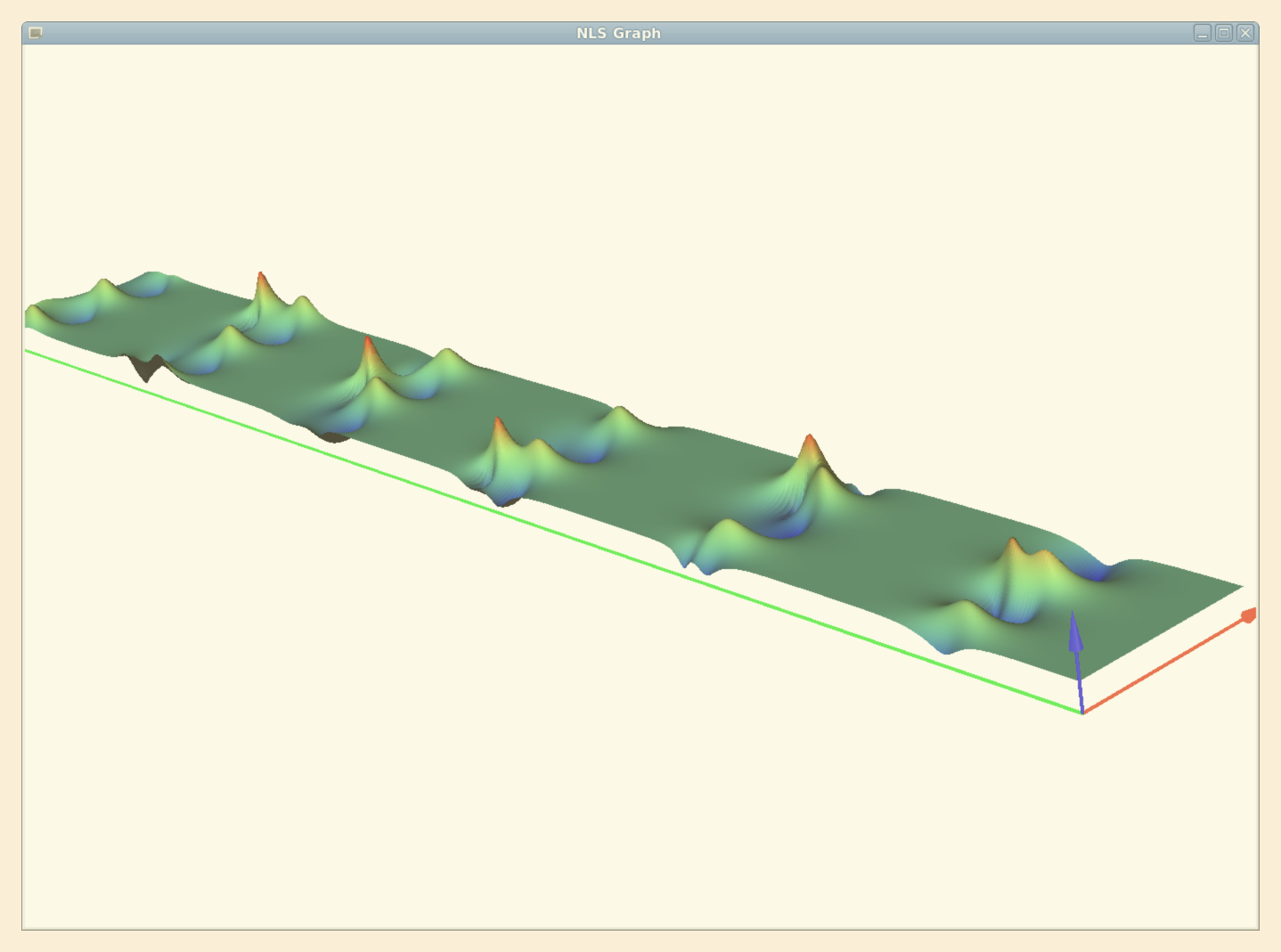} \ \includegraphics[width=5.15cm]{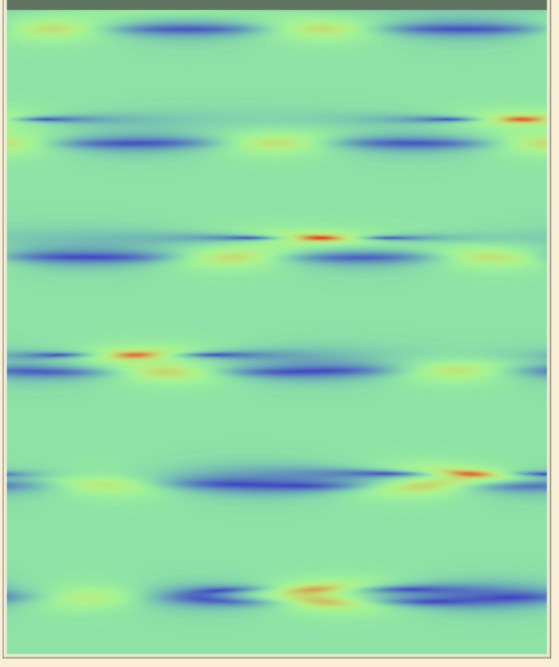}
\end{center}
Figures 2. The 3D and level plots of $|u(x,t)|$, obtained through the numerical integration of NLS via the SSFM, and generated by the data $L=7.05,~\eps=0.001,~c_1=(0.25+0.35 i)\eps$, $c_{-1}=(-0.15-0.2 i)\eps$, $c_2=(0.15-0.2 i)\eps$, $c_{-2}=(0.15+0.2 i)\eps$, in very good agreement with the theoretical formulas (\ref{sequence2RW}),(\ref{sequence_param}). For this choice of the data, the time differences $|T_1(|\alpha_1|)-T_2(|\alpha_2|)|=0.17$ and $|T_1(|\beta_1|)-T_2(|\beta_2|)|=0.15$ are small; therefore the two unstable modes are fully developed at almost the same times during the recursion, and have a truly nonlinear interaction for a longtime. The short axis is the $x$-axis, with $x\in [-L/2,L/2]$.

Using (\ref{time_asympt_2}), one obtains, as for $N=1$, {\it the uniform in space-time representation of the solution of the Cauchy problem, for $N=2$, in terms of elementary functions, describing the first $n$ RW appearances, for $0\le t\le (t_1^{(n)}+t_2^{(n)})/2+O(1)$ and $x\in [0,L]$: 
\beq\label{unif_sol_Cauchy_2}
\ba{l}
u(x,t)=\sum\limits_{m=0}^n A_2\Big(x,t;\phi_1,\phi_2,x_1^{(m)},x_2^{(m)},t_1^{(m)},t_2^{(m)},\rho^{(m)} \Big) -\frac{1-e^{4in(\phi_1+\phi_2)}}{1-e^{4i(\phi_1+\phi_2)}}e^{2it}, 
\ea
\eeq
where the parameters $x_j^{(m)},~ t_j^{(m)},~\rho^{(m)},~m\ge 0,~j=1,2$, are defined in (\ref{sequence_param}). Again this representation is accurate up to $O(\eps^2|\log\eps|)$ in the linear regions, and up to $O(\eps)$ in the nonlinear regions. 
}

We first remark that the above formulas (\ref{sequence2RW}),(\ref{sequence_param}),(\ref{unif_sol_Cauchy_2}) well describe the RW recurrence if $|t_1^{(1)}-t_2^{(1)}|\ll 1$, like in the numerical experiment in Figures 2, implying that the two nonlinear modes appear almost at the same time, and keep interacting for many recurrences. Instead, if the initial data are such that $|t_1^{(1)}-t_2^{(1)}|=O(1)$,  then the two modes tend to separate, and the slower one will start soon interacting with the faster one of the next generation; at this stage, the above formulas should be properly rewritten (we shall discuss this interesting issue in a subsequent paper).
 
We also remark that, for $N>1$, there is no time periodicity. 

At last, if the initial condition is the generic one (\ref{Cauchy}),(\ref{Fourier}), for $N=2$, then the solution (\ref{reg1_N UM}) contains also $O(\eps)$ oscillations coming from the excitation of the stable modes. But the behavior of the solution in the overlapping regions $1\ll |t|\ll O(|\log\eps|)$, for $t>0$ and $t<0$ is still given respectively by equations (\ref{overlapping2}) and (\ref{overlapping2b}), and the matching at $O(1)$ is not affected. Therefore the sequence of RWs is still described by equations (\ref{sequence2RW}), (\ref{sequence_param}), and the differences between the two Cauchy problems are hidden in the $O(\eps)$ corrections. 
\vskip 5pt
\noindent
{\bf The RW inverse problem}. We have established that, as far as the $O(1)$ RW recurrence is concerned, only the ``unstable part'' 
 $\eps_{unst}(x)\equiv \sum\limits_{j=1}^N (c_j e^{i k_j x}+c_{-j} e^{-i k_j x})$ of the initial perturbation (\ref{Fourier}) is relevant. Here we show how to reconstruct, from the relevant $O(1)$ data of the RW recurrence, the unstable part $\eps_{unst}(x)$ of the $O(\eps)$ initial perturbation, and we illustrate the procedure in the simplest case of $N=1$. 

If the measured period $L$ of the sequence of RWs is such that $\pi <L<2\pi$, then we are in the case $N=1$, with $k_1=\frac{2\pi}{L}, \ \phi_1=\arccos(\pi/L), \ \sigma_1=2\sin(2\phi_1)$. From the ``experimental'' observation of the points $(x_1^{(1)},t_1^{(1)})$ and $(x_1^{(2)},t_1^{(2)})$ at which the modulus of the first and respectively the second RW of the sequence have their maxima, we construct, from (\ref{parameters_1n}),(\ref{period}),(\ref{def_T1}),(\ref{def_alpha_beta}), the $O(\eps)$ parameters $\alpha_1,\beta_1$, through the formulas
\beq
\ba{ll}
|\alpha_1|=\frac{\sigma_1^2}{2}e^{-\sigma_1 t_1^{(1)}}, & \arg\alpha_1 = k_1 x_1^{(1)}+\phi_1-\frac{\pi}{2}, \\
|\beta_1|=\frac{\sigma_1^2}{2}e^{-\sigma_1(t_1^{(2)}-2 t_1^{(1)})}, & \arg\beta_1 = k_1 (x_1^{(2)}-2 x_1^{(1)})-\phi_1+\frac{\pi}{2}. 
\ea
\eeq
At last, from the knowledge of $\alpha_1,\beta_1$, one finally reconstructs the $O(\eps)$ Fourier coefficients $c_1,c_{-1}$ using (\ref{def_alpha_beta}):
\beq
c_1=\frac{\bar{\alpha_1}+e^{-2i\phi_1}\beta_1}{1-e^{-4i\phi_1}}, \ \ \ \ c_{-1}=\frac{e^{2i\phi_1}\alpha_1+\bar{\beta_1}}{1-e^{4i\phi_1}}.
\eeq  

\noindent
{\bf Conclusions and open problems}. We end this paper remarking that the analytic results obtained here for the Cauchy problem (\ref{Cauchy}), in the case of one and two unstable modes only, allow one to easily predict what is going to happen to the generic Cauchy problem (\ref{Cauchy}) in the case of $N$ unstable modes. One expects that 
{\it the solution describe and exact deterministic alternate recurrence of linear and nonlinear stages of MI, and that the nonlinear RW stages be described by the N-breather solution of Akhmediev type, depending on the $3N+1$ parameters
\beq
\{\phi_j,x^{(n)}_j,t^{(n)}_j,\rho^{(n+1)}\}_{j=1,\dots,N} ,
\eeq
all expressed in terms of the initial data via elementary functions. Therefore the RW dynamics is reduced to the study of the mapping 
\beq
\{\phi_j,x^{(n)}_j,t^{(n)}_j,\rho^{(n)}\}_{j=1,\dots,N} \ \to \ \{\phi_j,x^{(n+1)}_j,t^{(n+1)}_j,\rho^{(n+1)}\}_{j=1,\dots,N}
\eeq  
between the parameters of two consecutive RWs of the sequence.} This problem, solved in this paper for $N=1,2$, is presently under investigation by the authors using the same approach we proposed here, and no conceptual problems are expected to arise \cite{GS4}.  

Other interesting open problems are also under investigation, within the Cauchy problem (\ref{Cauchy}): a) the case $N\gg 1$ and the continuous limit of this theory; b) the case in which the arbitrary complex parameters $c_j$ appearing in the initial data are random; c) the effect of numerical and physical perturbations of the NLS model on the above RW recurrence; d) testing the above results on the RW recurrence of the NLS model in real experiments.
\vskip 5pt
\noindent
{\bf Acknowledgments}. Two visits of P. G. Grinevich to Roma were supported by the University of Roma ``La Sapienza'', and by the INFN, Sezione di Roma. P. G. Grinevich and P. M. Santini acknowledge the warm hospitality and the local support of CIC, Cuernavaca, Mexico, in December 2016, when part of this research was made. P.G. Grinevich was also partially supported by RFBR grant 17-51-150001.

We are grateful to M. Sommacal for introducing us to the Split Step Fourier Method and for showing us his personalized MatLab code. We acknowledge useful discussions with F. Briscese, F. Calogero, C. Conti, E. DelRe, A. Degasperis, A. Gelash, I. Krichever, A. Its, S. Lombardo, A. Mikhailov, D. Pierangeli, M. Sommacal and V. Zakharov.

\end{document}